\documentclass[sigconf]{acmart}

\usepackage{threeparttable}
\usepackage{comment}
\usepackage{enumitem}

\usepackage{algorithm}
\usepackage{algpseudocode}
\usepackage{balance} 
\copyrightyear{2024}
\acmYear{2024}
\setcopyright{acmlicensed}
\acmConference[WWW '24 Companion] {Companion Proceedings of the ACM Web Conference 2024}{May 13--17, 2024}{Singapore, Singapore.}
\acmBooktitle{Companion Proceedings of the ACM Web Conference 2024 (WWW '24 Companion), May 13--17, 2024, Singapore, Singapore}
\acmISBN{979-8-4007-0172-6/24/05} 
\acmDOI{10.1145/3589335.3651546}
\acmPrice{}



\begin{document}

\title{Position Bias Estimation with Item Embedding for Sparse Dataset}

%\author[1]{Shion Ishikawa, Yun Ching Liu, Young-joo Chung and Yu Hirate}
%\affil[1]{Address of first author}

\begin{comment}
    \author{
Shion Ishikawa$^*$,\hspace{0.7em} 
Yun Ching Liul$^*$,\hspace{0.7em}
Young-joo Chung$^*$ ,\hspace{0.7em}
Yu Hirate$^*$}

\affiliation{%
\vspace{0.6em} \institution{$^*$Rakuten Institute of Technology}
}

\affiliation{%
\vspace{0.6em}
  \institution{
  \{shion.ishikawa,\hspace{0.2em}
    yunching.liu,\hspace{0.2em}
    youngjoo.chung,\hspace{0.2em}
    yu.hirate\}@rakuten.com}
\vspace{0.6em}
}
\end{comment}

\settopmatter{authorsperrow=4}

\author{Shion Ishikawa}
\email{shion.ishikawa@rakuten.com}
\affiliation{%
  \institution{Rakuten Group, Inc.}
  \state{Tokyo}
  \country{Japan}
}
\author{Yun Ching Liu}
\email{yunching.liu@rakuten.com}
\affiliation{%
  \institution{Rakuten Group, Inc.}
  \state{Tokyo}
  \country{Japan}
}
\author{Young-joo Chung}
\email{youngjoo.chung@rakuten.com}
\affiliation{%
  \institution{Rakuten Group, Inc.}
%  \city{San Mateo}
  \state{California}
  \country{USA}
}
\author{Yu Hirate}
\email{yu.hirate@rakuten.com}
\affiliation{%
  \institution{Rakuten Group, Inc.}
  \state{Tokyo}
  \country{Japan}
}

%\renewcommand\Authands{ and }
%% "authornote" and 

\date{}
\begin{abstract}
Estimating position bias is a well-known challenge in Learning to Rank (L2R).  Click data in e-commerce applications, such as targeted advertisements and search engines, provides implicit but abundant feedback to improve personalized rankings. However, click data inherently includes various biases like position bias. Based on the position-based click model, Result Randomization and Regression Expectation-Maximization algorithm (REM) have been proposed to estimate position bias, but they require various paired observations of (item, position). In real-world scenarios of advertising, marketers frequently display advertisements in a fixed pre-determined order, which creates difficulties in estimation due to the limited availability of various pairs in the training data, resulting in a sparse dataset.  
We propose a variant of the REM that utilizes item embeddings to alleviate the sparsity of (item, position). Using a public dataset and internal carousel advertisement click dataset, we empirically show that item embedding with Latent Semantic Indexing (LSI) and Variational Auto-Encoder (VAE) improves the accuracy of position bias estimation and the estimated position bias enhances Learning to Rank performance. We also show that LSI is more effective as an embedding creation method for position bias estimation.

\end{abstract}

%\settopmatter{printacmref=false}
\maketitle

%% \renewcommand{\shortauthors}{Trovato and Tobin, et al.}





\section{Introduction}
Learning to Rank (L2R) \cite{cao2007learning} is a valuable approach to improving ranking performance in search engines, recommender systems \cite{yuan2017boostfm} \cite{rendle2009learning}, and targeted advertisement \cite{aharon2019carousel}.
Click data from rankings applications is a promising source for enhancing personalized rankings. Unlike explicit feedback, such as user ratings, click data reflects users' interaction with applications, making it a valuable form of \emph{implicit feedback}. However, due to its implicit nature, click data contains various biases, including position bias \cite{joachims2017accurately}. Position bias implies that the location of content influences its likelihood of being selected by users. To extract reliable signals from users' implicit feedback, previous research has focused on removing bias, in other words, denoising click data.

Click-Modeling is one approach to addressing the aforementioned issues. For instance, the position-based click model (PBM) \cite{richardson2007predicting} represents the probability of a user clicking on an item as the product of two probabilities: the probability of a user examining a specific position (position bias), and the probability of a user clicking on an item after examining it (item-user relevance).

In this study, we focus on estimating position bias under the PBM approach. While the cascade model  \cite{craswell2008experimental} has been extensively studied, it assumes a linear order of examining items and may not be applicable in all industrial scenarios where the order of item examination is not linear or follows a specific pattern, such as in the case of carousel ads with navigation buttons. For example, in the case of carousel ads with navigation buttons, when users click the left arrow button, the displayed ads shift to the left, causing users to examine the ads in a right-to-left order. Conversely, if they click the right arrow button, the order of examination will be reversed, with users examining the ads from left to right.

 The accuracy of position bias estimation is critical for the performance of recommender systems, because an inaccurate estimation will directly lead to a biased evaluation of item-user relevance.
 We have implemented the regression EM algorithm \cite{wang2018position} in the production carousel advertisement to estimate position bias from logged click data, without affecting real user traffic. Unfortunately, we discovered that this model didn't work well for this application due to \emph{the training data sparsity in the position bias estimation}.

%%%Selection bias refers to the bias that arises when the sample or data used for analysis is not representative of the users' behaviour. Selection bias affects not only the estimation of the average treatment effect \cite{hughes2019selection} but also regression models such as logistic regression \cite{mccullagh2008sampling}. With the biased input data, regression will fail to learn the map between the input to outcomes. Since the regression EM algorithm internally utilizes a regression model, selection bias can also affect the accuracy of position bias estimation.
The regression EM algorithm takes the tuple (item, position) as input data. Ideally, the items should be evenly distributed across all positions with a sufficient number of samples. \emph{However, in real-world scenarios, marketers often display advertisements in a fixed pre-determined order, resulting in most items being placed in only a few positions.} Moreover, in many cases, we observe one item is consistently placed in a single position. When we have 15 ads in our e-commerce platform, It means that we observe only $6.7\% (= 15/15^2)$ of total combinations in (ad, position) tuples. The position bias estimation suffers from it.

%Another example can be seen in search rankings, where the result of search ranking across many pages and a large number of positions exist. In this case, we observe each item is placed on only a small subset of positions across multiple pages.

In this study, we propose a method to address this sparsity issue. Our approach utilizes item similarities to estimate missing values in the (item, position) matrix. Specifically, when the (item $X$, position 1) entry is missing, we estimate its position using a similar item $X'$. To achieve this, we employ embedding techniques to create a representation that preserves similarity between items.
%Given embedding representation, we create a function that maps (item, feature) $n \times l$ matrix into (embedding vector, feature) $m \times l$ matrix where $m << n$, and represents the original sparse dataset with the dense embedding vectors. 
%From the item embedding, our study calculates the probability of each embedding vector given items. We then represent the original sparse matrix with a dense (embedding vector, position) matrix.
% To our best knowledge, this is the first paper to utilize item embeddings against the sparsity issue in position bias estimation. 
In our investigation, we discovered that item embeddings for estimating position bias have not been tested extensively with well-known tabular embedding baseline models like Latent Semantic Indexing (LSI) \cite{rosario2000latent} and Variational Autoencoder (VAE) \cite{kingma2019introduction}. Therefore, we aim to leverage these two models as an initial proposal in this study.

In summary, our study makes the following key contributions: 
\begin{itemize}[leftmargin=20pt]
  \item We introduce the novel problem of data sparsity in  (item, position) pairs for position bias estimation in a real-world scenario.
  \item We propose a novel method\footnote{We plan to publish code on https://github.com/rakutentech} for estimating position bias using Latent Semantic Indexing and Variational auto-encoder.
 \item We conduct extensive experiments on public and internal datasets, which demonstrates that our method, REM with item embedding, estimates position biases more accurately and improves learning-to-rank performance. 
 \item We find that LSI outperforms VAE in terms of accuracy of position bias estimation and personalized ranking.
 \end{itemize}

\section{Problem Setting}
\subsection{Position-based click model}
Let $i \in \mathcal I$ be an \textit{item} to be recommended (e.g., an advertisement or a product). Let $C \in \mathbb{R} _ {\geqslant 0}$ denote a reward variable (e.g., whether the displayed item was clicked or not). Let $\mathcal U$ denote a set of \textit{contexts}, such as a user's demographic features (e.g., age and occupation). Let $k$ be a position at which an item is displayed. 
The position-based click model assumes that the probability of clicks $P(C=1| i, u, k)$ can be represented as a product of two latent probabilities, as follows:
\begin{equation}
\label{def-positionbasedmodel}
    P(C=1| i, u, k) = P(R=1|i, u)P(E=1|k)
\end{equation}
where $P(E=1|k)$ represents the probability that position $k$ is examined by a user and $P(R=1|i, u)$ represents the probability that the item is related to the user. In short, the position-based click model assumes that if a user examines a particular position and the item in the position is relevant, they will click on the item. This model also assumes the examination process solely depends on a position and does not depend on other items. For simplicity, we denote relevance as $\mu(i, u) = P(R=1|i, u)$ and position bias as $\theta_k = P(E=1|k)$.

\subsection{Data sparsity in the position-based click model}
In the context of the position-based click model presented in Eq \ref{def-positionbasedmodel}, let $\mathcal D := \{(u_j, i_j, r_j, k_j)\}^n_{j=1}$ be a logged dataset of $n$ observations. Let $\mathcal{A}$ = $\{(i, k)\}$ be a set of actions containing all possible pairs of $(i, k)$.
We define a function $\pi : \mathcal{U} \to \Delta(\mathcal{A})$ as a policy, which maps each context $u \in \mathcal{U}$ to a distribution over the actions, where $\pi(i, k)$ represents the probability of assigning item $i$ to position $k$. $\Delta(\mathcal{A})$ is a simplex which satisfies $\Delta(\mathcal{A}) = \{a \in \mathbb{R}^K | \sum_{j=0}^K a_{j}=1, a_j \geq 0 \}$ where $K$ is a number of actions. The policy that creates the logging dataset $D$ is referred to as the logging policy.

In most cases, the logging policy is often determined by marketing experts based on their experience and expertise. As a result, it is mostly deterministic and static. As a result, the logged dataset contains a limited variation of $(i, k)$ pairs, which can lead to data sparsity that affects the estimation precision of position bias.

\section{Method}
\subsection{Item embeddings and modified representation of position-based click model}
As we discussed in the above sections, the sparsity problem exists for the $(i, k)$ pairs in the dataset $\mathcal D$. Hence, the estimation of position bias could be inaccurate due to spare and skewed data distribution. To mitigate this issue, we represent an item with latent embedding vectors $e \in \mathcal E$. Formally, we create a function that maps the $n \times l$ (item, feature) matrix into (item, embedding vector) $n \times m$ matrix where $m << l$. With the obtained embeddings, we then transform a sparse tuple $(i, k)$ into a dense tuple $(e, k)$. We first introduce the probability of an embedding vector given an item $P(e|i)$. Then we represent the probability of item-user relevance $P(R = 1|i, u)$ with embedding vectors, as follows.

\begin{equation}
\label{reward-embedding}
    P(R=1|i, u) = \sum_{e\in \mathcal E} P(e|i)P(R=1|e, u) 
\end{equation}
where $\sum_{e\in \mathcal E} P(e|i) = 1$. Based on $P(e|i)$ and Eq \ref{reward-embedding}, we sample partial reward $w$ from click $C$ based on the following equation.

\begin{equation}
\label{convert-rtow}
    P(w=1| e, u, k) = P(e|i)P(C=1| i, u, k)
\end{equation}

Then we can rewrite dataset $\mathcal D = \{(u_j, i_j, c_j, k_j)\}^n_{j=1}$ as $\mathcal D_e := \{(u_j, e_j, w_{j}, k_j)\}^{n|\mathcal E|}_{j=1}$.
Based on $\mathcal D_e$, we can rewrite the position-based click model in Eq. $\ref{def-positionbasedmodel}$ with embedding vectors.

\begin{equation}
\label{def-positionbasedmodel-embed}
    P(w=1| e, u, k) = P(R=1|e, u)P(E=1|k)
\end{equation}

\begin{table}
\caption{A toy example illustrating the benefit of item embedding against a sparse dataset.
Left table: logging policy (the distribution of item and their position assignment) which induces sparsity. It can be seen that item $i_0$ is only shown in position $k_0$.
Right table: Assignment of embedding vector to each item.
Bottom table: the resampled dataset with a new logging policy $\pi(e_l, k_l) = \sum_{j} p(e_l|i_j)\pi(i_j, k_l)$, showcasing reduced sparsity through item embeddings.}
\label{table:toyexample}
\begin{tabular}{c|ccc} 
\hline
& $\pi_b(i, k_0)$ & $\pi_b(i, k_1)$ & $\pi_b(i, k_2)$ \\
\hline
$i_0$ & 1/3 & 0 & 0 \\ 
$i_1$ & 0 & 1/3 & 0 \\ 
$i_2$ & 0 & 0 & 1/3 \\ 
\hline
\end{tabular}
\quad
\begin{tabular}{c|cc} 
\hline
& $p(e_0 | i)$ & $p(e_1 | i)$ \\
\hline
$i_0$ & 1/2 & 1/2  \\ 
$i_1$ & 1/3 & 2/3  \\ 
$i_2$ & 1/4 & 3/4  \\ 
\hline
\end{tabular}
\quad
\begin{tabular}{c|ccc} 
\hline
& $\pi(e, k_0)$ & $\pi(e, k_1)$ & $\pi(e, k_2)$ \\
\hline
$e_0$ & 1/6 & 1/9 & 1/12  \\ 
$e_1$ & 1/6 & 2/9 & 3/12  \\ 
\hline
\end{tabular}

\end{table}

Table \ref{table:toyexample} shows a toy example in which the sparsity of the original dataset is mitigated through item embedding. 

\subsection{Calculating the probability of embedding vector given item}
To use embeddings for position-based click models, we need to convert $\mathcal E$ into the probability space $[0, 1]$ to represent $p(e|i)$. 
We utilize the soft-max function to represent the conditional probability of embeddings with a given item, as introduced by Saito et al. \cite{saito2022off}. However, it is important to note that while Saito's work focused on building an off-policy evaluation with action embeddings for a large number of actions, our work addresses the issue of accurately estimating position bias in a sparse dataset. We define $p(e|i)$ as shown below:

\begin{equation}
\label{assignment_embedding}
     p(e|i) = \prod_{j=1}^m \frac{\exp(e_{i, j})}{\sum_{j'=1}^{m} \exp(e_{i, j'})}
\end{equation}

where $e_{i, j} \in \mathcal E$ refers to an element obtained from the embedding matrix (item $\times$ embedding). Here, we assume that when two items (e.g., item $i$ and $i'$) are similar, their corresponding representations in embedding space ($e_{i, j}$ and $e_{i', j}$) will be similar, and which leads to the similar conditional probabilities of embedding with a given item ($p(e|i)$ and $p(e|i')$). 
%This technique is particularly beneficial when dealing with sparse datasets. For instance, if we observe an item in only a specific position such as $(i_0, k_0)$ and $(i_1, k_1)$, and this is the case of the sparse dataset.
%In this case, we can obtain $p(e_0|i_0)$, $p(e_1|i_0)$, $p(e_0|i_1)$, and $p(e_1|i_1)$ through LSI and soft-max conversion. With a new representation of the dataset with tuples of (embedding vector, position), we can now have information for both positions ($(e_0, k_0)$ and $(e_0, k_1)$) for each embedding vector, leading to alleviate the problem of data sparsity as described in the previous section.
%TODO: Add toy example of position bias estimation and embedding
\subsection{Regression EM algorithm with embedding vectors}
Our goal is to obtain an accurate position bias $\theta_k$. Under the position-based click model in Eq. \ref{def-positionbasedmodel}, we apply the Regression EM algorithm \cite{wang2018position} to optimize $P(R=1|i, u)$ and $P(E=1|k)$ from the observed $P(C=1| i, u, k)$. 

We summarize our Regression EM algorithm with embedding in Algorithm 1. At iteration $t + 1$, the Expectation step will update the distribution of hidden variable $E$ and $R$ from $\theta_k^{(t)}$ and $\mu^{(t)}(i, u)$. We use the Gradient Boosted Decision Tree (GBDT) method to learn the function $\mu(u, e)$.

\begin{align*}
\label{hidden_distributions}
    P(E=1, R=1| w=1, e, k) &= 1 \\
    P(E=1, R=0| w=0, e, k) &= \frac{\theta_k^{(t)}(1-\mu^{(t)}(e, u))}{1-\theta_k^{(t)}\mu^{(t)}(e, u)} \\
    P(E=0, R=1| w=0, e, k) &= \frac{(1-\theta_k^{(t)})\mu^{(t)}(e, u)}{1-\theta_k^{(t)}\mu^{(t)}(e, u)} \\
    P(E=0, R=0| w=0, e, k) &= \frac{(1-\theta_k^{(t)})(1-\mu^{(t)}(e, u))}{1-\theta_k^{(t)}\mu^{(t)}(e, u)}
     \stepcounter{equation}\tag{\theequation} 
\end{align*}

The Maximization step calculates $\theta_k^{(t+1)}$ and $\mu^{(t+1)}(e, u)$ using the probabilities from the Expectation step.
 \begin{align*}
 \label{mamixization_step}
     \theta_k^{(t+1)} = \frac{\sum_{(u, e, w, k')\in \mathcal D_e} \mathcal{I}_{k'=k}(w+(1-w)P(E=1|u, e, w, k))}{\sum_{(u, e, w, k')\in \mathcal D_e} \mathcal{I}_{k'=k}} \\
     \mu^{(t+1)}(e, u) = \frac{\sum_{(u, e', w, k)\in \mathcal D_e} \mathcal{I}_{e'=e}(w+(1-w)P(R=1|u, e, w, k))}{\sum_{(u, e', w, k)\in \mathcal D_e} \mathcal{I}_{e'=e}}
      \stepcounter{equation}\tag{\theequation} 
 \end{align*}
where $\mathcal{I}$ shows the indicator function.

\begin{algorithm}
\caption{Regresssion EM algorithm with embedding}\label{alg:regression-em-embed}
\begin{algorithmic}
\Require logging dataset $\mathcal D = \{($user's context $u,$ item $i,$ click $c,$ position $k)\}$, position bias: $\{\theta_k\}$, relevance between embedding $e$ and user's context $u$: $\mu(u, e)$, assignment of embedding to item: $P(e|i)$
\ForAll {$(u, i, c, k) \in \mathcal D$} 
    \State Sample $w \in \{0, 1\}$ from $c$ with $P(e|i)$ based on Eq. \ref{convert-rtow}.
\EndFor
\State Prepare $\mathcal D_e = \{(u, e, w, k)\}$ from sampled $\mathbf{w}$
\Repeat 
    \State Let $S = \{\}$
    \ForAll {$(u, e, w, k) \in \mathcal D_e$} 
        \State Sample $r \in \{0, 1\}$ from $P(R = 1|u, e, w, k)$ based on Eq. \ref{hidden_distributions}.
        \State $S = S \cup (u, e, r)$
    \EndFor
    \State Update $\mu(u, e) = GBDT(\mu(u, e), S)$ 
    \State Update $\{\theta_k\}$ based on Eq. \ref{mamixization_step}.
\Until {Convergence} \\
\Return{$\{\theta_k\}$, $\mu(e, u)$}
\end{algorithmic}
\end{algorithm}

\section{Experiments}

\begin{table*}[htbp]
%\begin{table}
\label{table:ranking_metric}
\centering
\renewcommand\thetable{3}
\caption{Impact on ranking metrics with improved position bias}
\begin{tabular}{c|ccc|ccc} 
\hline
  & \multicolumn{3}{|c}{Open Bandit Dataset} & \multicolumn{3}{|c}{Rakuten Ichiba dataset}  \\
 & MRR & MAP@5 & MAP@10
 & MRR & MAP@5 & MAP@10   \\
\hline
Regression EM (REM) & 8.773e-3 & 1.709e-3 & 8.773e-4  & 0.7179 & 0.1573 & 7.394e-2 \\ 
VAE + REM & 8.777e-3 & \textbf{1.712e-3} & 8.777e-4 & 0.7599 & 0.1583 &  7.673e-2  \\ 
LSI + REM & \textbf{9.049-3} & 1.708e-3 & \textbf{9.049-4} & \textbf{0.7716} & \textbf{0.1613} & \textbf{7.795e-2}  \\ 
\hline
\end{tabular}
\end{table*}

\subsection{Experiments setup}
In this section, we conduct extensive experiments to answer the following questions.
\begin{itemize}[leftmargin=20pt]
  \item \textbf{(RQ1)}: How accurately does the proposed method estimate the position bias from a sparse dataset?
  \item \textbf{(RQ2)}: How does the proposed method perform in a real-world application?
\end{itemize}

\subsubsection{Dataset}
We evaluate the proposed method with following two datasets:
\begin{itemize}[leftmargin=20pt]
  \item Open Bandit Dataset \cite{saito2020open}: It is a large public dataset, which contains 1,374,327 instances with 88 features and 80 items. The dataset has two logging policies: a uniform random and Thompson Sampling. It has only three positions which users see simultaneously and the difference among trained position biases was less than 0.1\%.
  Therefore, we instead introduce 10 virtual positions and resample clicks. We follow the procedure in \cite{Ren_2022} to sample clicks. In the PBM model (Eq \ref{def-positionbasedmodel}), we replace $P(E = 1 |k)$ with position biases from eBay's experiment \cite{Aslanyan_2019}, and we resample clicks as $click_{i, u, k} \thicksim U(P(C=1 |i, u, k))$. 
  \item Rakuten Ichiba Dataset: Rakuten is one of the largest e-commerce companies in Japan. This platform has a widget of carousel ads containing 15 ads, with three ads displayed simultaneously. From the four-day access log of this widget, we obtained 494,329 instances with 156 features. As an A/B test, one logging policy was determined heuristically by marketers, leading to sparse pairs of (item, position). The other policy was a uniform random as control group.
\end{itemize}
\subsection{Effect on position bias estimation on sparse datasets (RQ1)}
To evaluate the position bias estimation against a sparse datasets, we constructed a sparse dataset by restricting an item to being assigned to a single position from a dataset with a uniform random logging policy.
We applied the proposed method to this sparse dataset and assessed its ability to recall the position bias of the original dataset in terms of RMSE. But in Open Bandit Dataset, because we know the true values of the virtual position biases, we compare our trained values with them.

\begin{figure}[htbp]
  \begin{minipage}[b]{1.0\linewidth}
    \centering
    \mbox{\raisebox{0mm}{\includegraphics[keepaspectratio, scale=0.45]{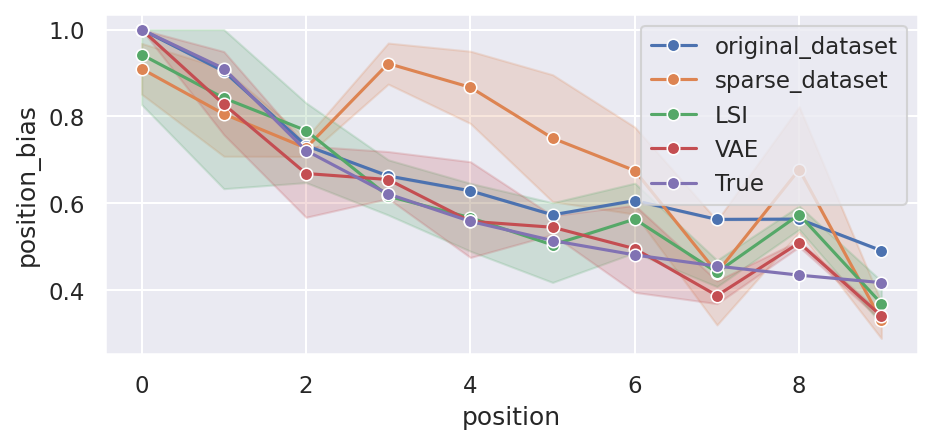}}}  
  \end{minipage}
  \begin{minipage}[b]{1.0\linewidth}
    \centering
    \includegraphics[keepaspectratio, scale=0.45]{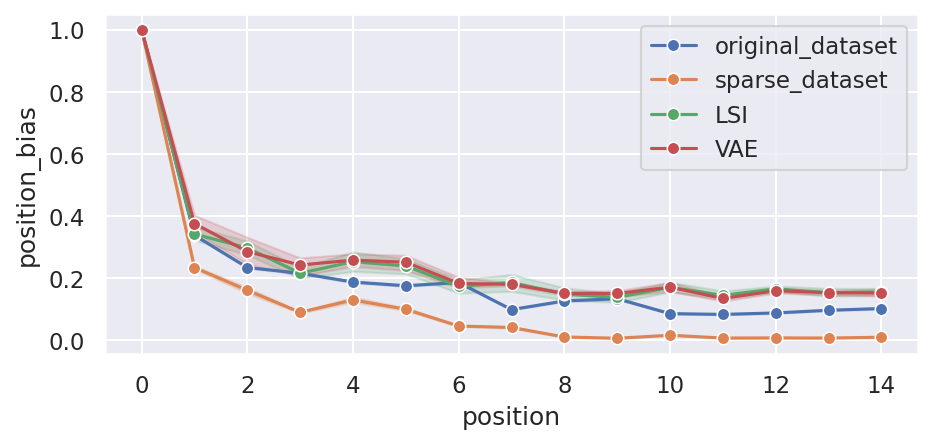}
  \end{minipage}

   %  \Description[(Left) The line plot of position bias at each position. (Right) The barplot of RMSE for each policy. VAE improved RMSE relatively by 10.3\% and LSI improved RMSE relatively by 33.4\%]

  \caption{The comparison of position bias estimation in Open Bandit Dataset (top) and Rakuten Ichiba Dataset (bottom).}
  Lines show position biases estimated from different conditions. (Blue): original dataset with Regression EM. (Yellow): sparse dataset with Regression EM. (Green): sparse dataset with Latent Semantic Indexing. (Red): sparse dataset with Variational Auto-Encoder. (Violet): Virtual position biases from the experiment \cite{Aslanyan_2019}. We regard them as true position biases.
\label{fig_realdata}
\end{figure}
\begin{table}[htbp]
\label{table:rmse}
\renewcommand\thetable{2}
\caption{RMSE of position bias estimation for a sparse dataset}
\begin{tabular}{c|cc} 
\hline
  & Open Bandit Dataset & Rakuten Ichiba dataset  \\
\hline
REM & 0.271 $\pm$ 0.0277 & 0.292 $\pm$ 0.1138  \\ 
VAE + REM & 0.257 $\pm$ 0.1147 & 0.265 $\pm$ 0.1203  \\ 
LSI + REM & \textbf{0.219 $\pm$ 0.0989} & \textbf{0.260 $\pm$  0.1137}  \\ 
\hline
\end{tabular}
\end{table}
Fig.\ref{fig_realdata} and Table 2 show the performance when we applied the regression EM algorithm (REM) with Latent Semantic Index, REM with Variational Auto-Encoder, and vanilla REM to the sparse dataset. Table 2 shows the mean and standard deviation of RMSE with 5 trials. It can be seen that VAE improved RMSE relatively by 5.1\% and LSI improved RMSE relatively by 19.2\%. This result indicates the item embedding leads to the accurate estimation of position bias, and LSI outperforms VAE in terms of RMSE.

\subsection{Effects on Ranking recommendation (RQ2)}
We evaluate the performance of learned position bias estimation in the recommendation model. We first fixed position biases with the learned values from the previous experiment, and then we trained GDBT estimators under the PBM model. Table 3 shows the performance of mean reciprocal rank (MMR) and mean average precision (MAP) with the top 5 positions and top 10 positions. Values are the mean of 5 trials. This result indicates that item embedding improved ranking metrics, and except for MAP@5 in the Open Bandit Dataset, LSI performs better than VAE.

\section{Conclusion}
In this paper, we first introduced the data sparsity issue in the position bias estimation under the Position-based click model. To overcome this issue, we proposed the Regression EM algorithm with embedding (with LSI and VAE). By utilizing the assignment probability $p(e|i)$ obtained from the item embedding, we sampled various pairs of (embedding vector, position) and applied the Regression EM algorithm to the sampled dataset. With the two real-world datasets, we validated that our method outperforms the vanilla REM in terms of both position bias estimation and recommendation.

\bibliographystyle{www2024_short}
\balance
\bibliography{www2024_short}

\begin{comment}
\subsection{Appendix}
\subsubsection{Click simulation with virtual position in Open Bandit Dataset}.
Here we explain how we introduce virtual positions and generate clicks based on them. We employ the following position-based click model.
\begin{equation}
\label{def-positionbasedmodel}
    P(C=1| i, u, k) = P(R=1|i, u)P(E=1|k)
\end{equation}
$P(E=1|k)$ is position bias. Position bias is obtained from eye-tracking experiments or previous other study of position bias. $P(R=1|i, u)$ is the content relevance. When the dataset doesn't have the position column, the vector of clicks represents content relevance. If the dataset has an explicit rating (such as in MovieLens), we transform the explicit rating to the probability $P(R|i, u)$ \cite{Ren_2022}. Finally, we sample clicks uniformly from $P(C=1| i, u, k)$.
The previous study \cite{Ren_2022} obtained $P(R=1|i, u)$ from explicit rating and obtained $P(E=1|k)$ from eye-tracking experiment.
In our case, Open Bandit Dataset has three positions, however, positions are shown on the web page together and the trained $P(E=1|k)$ are almost the same. 
To make clicks dependent on the positions more, we instead regard $P(E=1|k)$ as position bias from eBay's previous research \cite{Aslanyan_2019} and regard content relevance as a click vector in the dataset, and then we resampled clicks from new $P(C=1| i, u, k)$.
\end{comment}

\end{document}